\documentclass[aps,twocolumn,showpacs,floatfix,superscriptaddress]{revtex4}
\usepackage{graphicx}
\usepackage{amsmath}
\usepackage{amsfonts}
\usepackage{amssymb}

\begin{document}




\title{Pfaffian-like ground state for 3-body-hard-core bosons in 1D lattices}

\author{B. Paredes}
\affiliation{Institut f\"{u}r Physik, Johannes Gutenberg-Universit\"{a}t, Staudingerweg 7, D-55099 Mainz, Germany}
\affiliation{Max--Planck Institut f\"{u}r Quantenoptik,
Hans-KopfermannStr. 1, Garching, D-85748 Germany}
\author{T. Keilmann}
\affiliation{Max--Planck Institut f\"{u}r Quantenoptik,
Hans-KopfermannStr. 1, Garching, D-85748 Germany}
\author{J. I. Cirac}
\affiliation{Max--Planck Institut f\"{u}r Quantenoptik,
Hans-KopfermannStr. 1, Garching, D-85748 Germany}

\begin{abstract}
We propose a Pfaffian-like Ansatz for the ground state of bosons subject to 3-body infinite repulsive interactions in a 1D lattice. Our Ansatz consists of the symmetrization over all possible ways of distributing the particles in two identical Tonks-Girardeau gases.  We support the quality of our Ansatz with numerical calculations and propose an experimental scheme based on mixtures of bosonic atoms and molecules in 1D optical lattices in which this Pfaffian-like state could be realized. Our findings may open the way for the creation of non-abelian anyons in 1D systems.
\end{abstract}

\date{\today}
 \maketitle

Beyond bosons and fermions, and even in contrast to the fascinating abelian anyons (AA) \cite{AA}, non-abelian anyons (NAA) \cite{NAA} exhibit an exotic statistical behavior: If two different exchanges are performed consecutively among identical NAA, the final state of the system will depend on the order in which the two exchanges were made.
NAA appeared first in the context of the fractional quantum Hall effect (FQHE) \cite{NAA}, as elementary excitations of exotic states like the Pfaffian state \cite{Pfaff, Cluster}, the exact ground state of quantum Hall Hamiltonians with 3-body contact interactions.
Recently, the possibility of a fault tolerant quantum computation based on NAA \cite{Fault_Tolerant} has boosted the investigation of new models containing NAA \cite{Models_NAA}, as well as the search for techniques for their detection and manipulation \cite{Detect_NAA}. Meanwhile, the versatile and highly controllable atomic gases in optical lattices \cite{Bloch} have opened a door to the near future implementation of those models as well as for the artificial creation of non-Abelian gauge potentials \cite{Zoller_gauges}.

All actual models containing NAA are 2D models. The motivation of the present work is the foreseen possibility of creating NAA in one-dimension (1D). This long-term goal requires in the first place to define the concept of NAA, which is essentially 2D, in 1D. For abelian anyons (AA) this generalization has been already made by Haldane \cite{Haldane_AA}. Within his generalized definition the spinon excitations of 1D Heisenberg antiferromagnets are classified as $\frac{1}{2}$-AA \cite{Haldane_AA}. This classification becomes very natural through the connection between the 1D antiferromagnetic ground state (for a long-range interaction model, the Haldane-Shastry model \cite{Haldane_Shastry}) and the Laughlin state \cite{Laughlin} for bosons at $\nu=1/2$. In a similar way we anticipate that a connection can be established between quantum Hall models containing NAA and certain long-range 1D spin models exhibiting NAA within a generalized definition \cite{Belén}.

Here, far from analyzing the above questions in general, our aim is to pave the way for the creation of exotic Pfaffian-like states in 1D systems, which we believe may serve as the basis to create NAA.
We present a realistic 1D system whose ground state is very close to a Pfaffian-like state. 
The actual system we consider is that of bosonic atoms in a 1D lattice with infinite repulsive 3-body on-site interactions, which we call 
{\em 3-hard-core bosons}. Inspired by the form of the fractional quantum Hall Pfaffian state for bosons \cite{Cluster, Gunn}, we propose an Ansatz for the ground state of our system. This Ansatz is a symmetrization over all possible ways of distributing the particles in two identical Tonks-Girardeau (T-G) gases \cite{Girardeau, nature}. Comparison of the Ansatz with numerical calculations for lattices up to 40 sites yields very good agreement. As for fractional quantum Hall systems, NAA may be created here by creating pairs of quasiholes,  each quasihole being in a different cluster \cite{Cluster}. This possibility will be discussed elsewhere \cite{Belén}.

Three-body collisions among single atoms rarely occur in nature. However, they can be effectively simulated by mixtures of bosonic particles and molecules. This has been proposed by Cooper \cite{Cooper} for a rapidly rotating gas of bosonic atoms and molecules. Here, we show that a system of atoms and molecules in a 1D lattice can in a similar way effectively model 3-hard-core bosons. 
We will show that the conditions to realize this situation lie within current experimental possibilities.

{\em $3$-hard-core bosons}.
We consider a system of bosonic atoms in a 1D lattice with repulsive 3-body on-site interactions. This system is described by
the Hamiltonian:

\begin{equation}
H=-t\sum_{\ell}( a^{\dagger}_\ell a^{\,}_{\ell+1}+h.c. )+ U_3\sum_\ell 
(a^{\dagger}_\ell )^3(a^{\,}_\ell)^3,
\label{Ham}
\end{equation}
where the operator $a^{\dagger}_\ell$ ($a_\ell$) creates (annihilates) a boson on site $\ell$, $t$ is the tunneling probability amplitude, and $U_3$ is the on-site interaction energy. 
From now on we will consider the limit $U_3 \to \infty$. In this limit the Hilbert space is projected onto the subspace of states with occupation numbers
$n_\ell=0,1,2$ per site. We will refer to bosons subject to this condition as 3-hard-core bosons. 
The projected Hamiltonian has the form
\begin{equation}
H_3=-t\sum_{\ell}  ( a^{\dagger}_{3,\ell} a^{\,}_{3,\ell+1}+h.c. ), \label{Ham_3_body}
\end{equation}
where the 3-hard-core bosonic operators $a_{3,\ell}$ 
obey $(a_{3, \ell})^3=0$  and satisfy the commutation relations 
$[a^{}_{3,\ell^{}},a^{\dagger}_{3,\ell^{\prime}}]=\delta_{\ell, \ell^{\prime}}\left(1-\frac{3}{2}(a^{\dagger}_{3,\ell})^2 (a^{\,\,}_{3,\ell})^2\right)$.
These operators can be represented by $3 \times 3$ matrices of the form
$a^{\dagger}_{3, \ell}= \left( \begin{array} {ccc}
0 & 1 & 0 \\
0 & 0 & \sqrt{2} \\
0 & 0 & 0
\end{array} \right)$.
In contrast to the usual hard-core bosonic operators \cite{Sachdev}, which are directly equivalent to spin-1/2 operators, the operators $a^{\dagger}_{3}$,  $a^{\,}_{3}$ are related to spin-1 operators $\{S^+,S^-,S^z\}$ in a non-linear way:
$a_{3} \rightarrow S^+\left(\frac{1}{\sqrt{2}}+(\frac{1}{\sqrt{2}}-1) S^z\right)$.
This mapping leads to a complicated equivalent spin Hamiltonian (with third and fourth order terms) which seems hard to solve.

In the following we present an Ansatz wave function for the ground state of Hamiltonian 
(\ref{Ham_3_body}). Our Ansatz is inspired by the form of the ground state for fractional quantum Hall bosons subject to three body interactions \cite{Pfaff, Cluster, Gunn}. The reason to believe that this inspiration may be good is 
the deep connection already demonstrated for the case of two-body interactions between ground states of certain 1D models and those of 2D particles in the lowest Landau level (LLL) \cite{Haldane_Shastry}. 

Let us now turn for a moment to the problem of bosons in the LLL subject to the 3-body interaction potential $\sum_{i\neq j \neq k} \delta^2 (z_i-z_j) \delta^2 (z_i-z_k)$ \cite{Pfaff}, with $z_i=x_i+iy_i$ being the complex coordinate in the 2D plane. For infinite interaction strength the exact ground state of the problem is the Pfaffian state \cite{Pfaff, Gunn}: 
\begin{equation}
\Phi_3\propto \mathcal{S}_{\uparrow, \downarrow} \left( \prod_{i<j}^{N/2}(z_i^{\uparrow}-z_j^{\uparrow})^2
\prod_{i<j}^{N/2}(z_i^{\downarrow}-z_j^{\downarrow})^2 \right).
\label{Cluster_state}
\end{equation}
This state is constructed in the following way. Particles are first arranged into two identical $\nu=1/2$ Laughlin states \cite{Laughlin}, $\Phi_2^{\sigma} \propto \prod_{i<j}^{N/2}(z_i^{\sigma}-z_j^{\sigma})^2$ labeled by $\sigma=\uparrow,\downarrow$. 
Then the operator $\mathcal{S}_{\uparrow, \downarrow}$ symmetrizes over the two ''virtual'' subsets of coordinates $\{z_i^{\uparrow}\}$ and $\{z_i^{\downarrow}\}$. Note that the Laughlin state 
$\Phi_2^{\sigma}$ of each cluster is a zero energy eigenstate of the 2-body interaction potential
$\sum_{i\neq j} \delta (z_i-z_j)$. This guarantees that in a state of the form (\ref{Cluster_state}) three particles can never coincide in the same position: for any trio, two of them will belong to the same group and cause the wave function $\Phi_3$ to vanish.

In direct analogy with equation (\ref{Cluster_state}) we propose the following Ansatz for the ground state of Hamiltonian (\ref{Ham_3_body}):

\begin{equation}
\Psi_3\propto{\mathcal{S}}_{\uparrow, \downarrow} \left( \prod_{i<j}^{N/2}\left|\sin (x_i^{\uparrow}-x_j^{\uparrow})\right|\prod_{i < j}^{N/2} \left|\sin (x_i^{\downarrow}-x_j^{\downarrow})\right|\right).
\label{Ansatz}
\end{equation}
This Ansatz has the same structure as (\ref{Cluster_state}), but the Laughlin state has been substituted by a Tonks-Girardeau (T-G) state \cite{Girardeau}, $\Psi_2^{\sigma}\propto
\prod_{i<j}^{N/2}\left|\sin (x_i^{\sigma}-x_j^{\sigma})\right|$, with $x_i^{\sigma}=2\pi/Mi$,
$i=1, \ldots, M$, $M$ being the number of lattice sites. This state is the ground state of hard-core 1D lattice bosons with Hamiltonian $H_{2, \sigma}=-t \sum_{\ell}( a^{\dagger}_{2,\sigma,\ell} a^{\,}_{2,\sigma,\ell+1}+h.c. )$ and periodic boundary conditions \cite{nature}. Here, $a^{\,}_{2, \sigma}$ are hard-core bosonic operators satisfying $(a^{\,}_{2, \sigma})^2=0$.
Written in second quantization the Ansatz (\ref{Ansatz}) takes the form:
$
|\Psi_3 \rangle =\mathcal{P} \left( |\Psi_2^{\uparrow}\rangle \otimes |\Psi_2^{\downarrow}\rangle \right)$,
where $\mathcal{P}$ is a local operator of the form $\mathcal{P}=\mathcal{P}_\ell^{\otimes M}$, and $\mathcal{P}_\ell$ is an operator mapping the single-site 4-dimensional Hilbert space of two species of hard-core bosons to the 3-dimensional one of 3-hard-core bosons (see Fig. 1).

\begin{figure}[h]
	\begin{center}
		\includegraphics[width=0.4\textwidth]{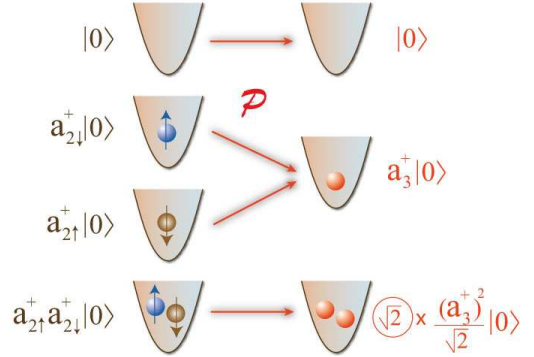}
	\end{center}
	\label{fig:mapping}
	\caption{Schematic representation of the operator $\mathcal{P}_\ell$ mapping the single-site 4-dimensional Hilbert space of two species of hard-core bosons to the 3-dimensional one of 3-hard-core bosons.} 
\end{figure}

Let us analyze different characteristic properties of our Ansatz. Taking into account the well known result for a T-G gas, namely the scaling of the one-particle correlation function 
$\langle a_{\ell+\Delta}^{\dagger} a_{\ell} \rangle$ as $\Delta^{-1/2}$ for large $\Delta$ \cite{nature}, we can derive the following asymptotic behavior for the one-body and two-body correlation functions for the Ansatz (\ref{Ansatz}):
\begin{eqnarray}
\langle a_{\ell+\Delta}^{\dagger} a_{\ell} \rangle & \longrightarrow & \Delta^{-1/4} \label{correlation}\\
\langle a_{\ell+\Delta}^{\dagger} a_{\ell+\Delta}^{\dagger} a_\ell a_\ell \rangle & \longrightarrow & 
\Delta^{-1}\label{correlation_doubly}.
\end{eqnarray}
The result (\ref{correlation_doubly}) can be easily derived by noticing that $\langle a_{\ell+\Delta}^{\dagger} a_{\ell+\Delta}^{\dagger} a_\ell a_\ell \rangle \propto 
\langle \Psi_2^{\uparrow} |a_{\ell+\Delta, \uparrow}^{\dagger} a_{\ell, \uparrow} | \Psi_2^{\uparrow} \rangle
\langle \Psi_2^{\downarrow} |a_{\ell+\Delta, \downarrow}^{\dagger} a_{\ell, \downarrow} | \Psi_2^{\downarrow} \rangle \rightarrow
\Delta^{-1/2}\Delta^{-1/2}$. The proof of (\ref{correlation}) is more involved and we will give just numerical evidence from our calculations bellow (see inset of Fig. 3).
The two-body correlation (\ref{correlation_doubly}) is indeed in our case the (one-particle) correlation function for on-site pairs. This means that whereas the system seen as a whole exhibits some kind of coherence (the spatial correlation decaying slowly as $\Delta^{-1/4}$) the underlying system of on-site pairs is in a much more disordered state (with a fast correlation decay as $\Delta^{-1}$).
This is in contrast to what happens in a weakly interacting bosonic gas in which coherence between sites is independent of their occupation number.
We can also obtain analytical expressions for the relative occupation of single and doubly occupied sites. The average number of doubly occupied sites is $n_2=\langle a_{\ell}^{\dagger} a_{\ell}^{\dagger} a_\ell a_\ell \rangle/2= 
\nu^2/2$, and the one of single occupied sites is given by $n_1=\langle a_{\ell}^{\dagger} a_{\ell}(2-n_\ell)\rangle=\nu(2-\nu)$. This distribution is very different from the Poissonian one, for which we have $n_2^{\textrm{Po}}/n_1^{\textrm{Po}}=\nu/2$.



As an additional property, the Ansatz (\ref{Ansatz}) has particle-hole symmetry. This means that for a filling factor of the form $\nu=N/M=2-\eta$, with $N$ the number of particles, the state we propose is just the Ansatz for holes at $\nu_h=\eta$. However, as we can clearly see from its matrix representation, the Hamiltonian (\ref{Ham_3_body}) does not exhibit this symmetry. This tells us that our Ansatz may not work, as we will see, for the whole regime of filling factors.

{\em Numerical calculations}.
In order to determine the quality of our Ansatz we have performed a numerical calculation for the ground state of Hamiltonian (\ref{Ham_3_body}). To obtain the numerical ground state $|\Psi_{\textrm{ex}} \rangle$ we have used variational Matrix Product States (MPS)  of the form $\sum_{s_1, \ldots, s_N=1}^d \mathrm{Tr}
\left( A_1^{s_1}\ldots A_N^{s_N} \right) \vert  s_1 \ldots s_N
\rangle$ \cite{MPS}, with matrices $A$ of dimension $D=12(15)$, and $d=3$.  We can estimate the error of this calculation to be smaller than $10^{-5}$ for the system sizes ($M\le40$) we have considered. To calculate the overlap of $|\Psi_{\textrm{ex}} \rangle$ with our Ansatz $|\Psi_3 \rangle$ we first construct the MPS state that best approximates $|\Psi_3 \rangle$ for a given $D$. This is done in the following way.
We first build the MPS ground state of Hamiltonian $H_2$.
We then take the tensor product of this state with itself, obtaining a MPS with $d=4$, which is closest to $|\Psi_2^{\uparrow}\rangle \otimes |\Psi_2^{\downarrow}\rangle$. The dimension of the matrices of this state is very large and we use a reduction algorithm \cite{Juanjo} to reduce it to the initial size. Finally we apply the operator $\mathcal{P}$ by local tensor contraction and normalize the resulting $d=3$ MPS state. For the matrix dimensions we used the  
error made was always smaller than $10^{-3}$.

\begin{figure}
	\begin{center}
		\includegraphics[width=0.4\textwidth]{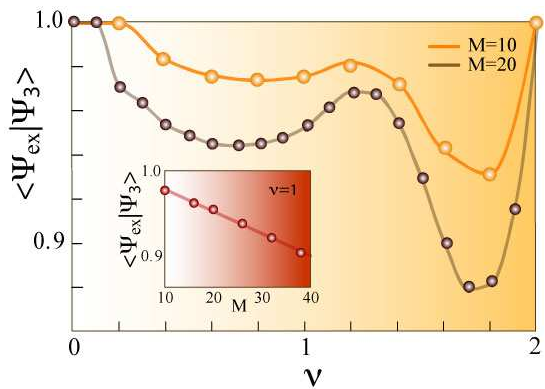}
	\end{center}
	\label{fig:overlap}
	\caption{Overlap $\langle \Psi_{\textrm{ex}} | \Psi_{3} \rangle$. The main plot shows the dependence of the overlap with the filling factor $\nu=N/M$ for a system of $M=10$ (orange circles) and $M=20$ (brown circles) lattice sites. The inset shows the decrease of the overlap with increasing system size $M$, at $\nu=1$.}
\end{figure}

The results are shown in Fig. 2. The main plot shows the overlap $\langle \Psi_{\textrm{ex}}|\Psi_3 \rangle$ as a function of the filling factor $\nu=N/M$ for a fixed system size. We find very good overlaps (0.98-0.96) for $\nu \le 1.25$. For  $\nu > 1.25$ the overlap decreases.
The inset shows the overlap as a function of increasing system size $M$, at fixed filling factor
$\nu=1$. At $M=38$, the maximum size we have considered numerically, the overlap is still good ($\approx 0.90$).

Fig. 3 shows the statistical distribution of doubly and single occupied sites for $|\Psi_{\textrm{ex}}\rangle$, which is very close to the one of the Ansatz $|\Psi_3\rangle$, and clearly different from the Poissonian distribution typical of a weakly interacting Bose gas.
Fig. 4 shows the momentum distribution for particles and on-site pairs together with the long-range scaling of their spatial correlation functions. We can clearly see how the exact state exhibits all characteristic behaviors that we have discussed above for the Ansatz.

{\em Experimental proposal}.
\begin{figure}[b]
	\begin{center}
		\includegraphics[width=0.4\textwidth]{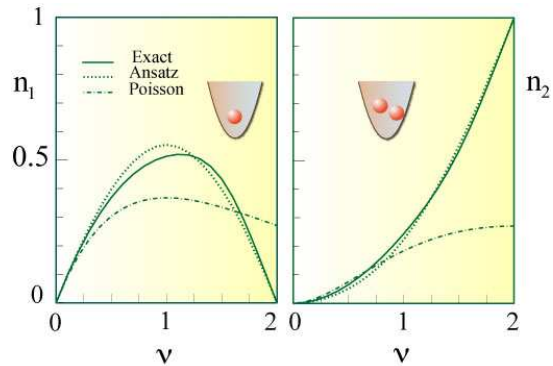}
	\end{center}
	\label{fig:ocupaciones}
	\caption{Average occupation of sites with one (left figure) and two particles (right figure) for the exact (solid lines), Ansatz (dotted lines) and Poissonian distribution (dot-dashed lines), as a function of the filling factor $\nu$. The system size is $M=20$.}
\end{figure}

\begin{figure}
	\begin{center}
		\includegraphics[width=0.45\textwidth]{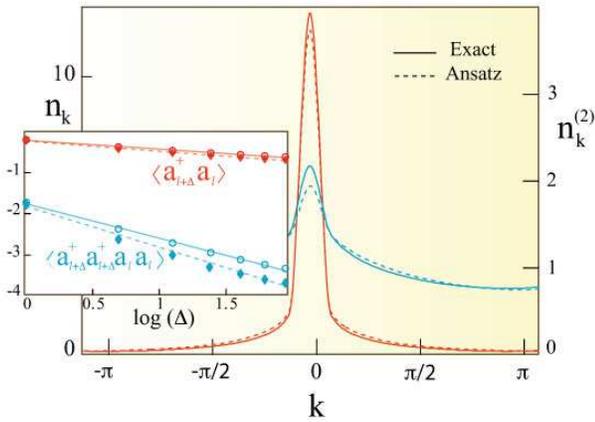}
	\end{center}
	\label{fig:momentum}
	\caption{Quasi-momentum distribution of particles $n_k \propto \sum_{\ell, \Delta} e^{-ik\Delta} \langle a^{\dagger}_{\ell+\Delta}a_\ell \rangle$ (orange, left axis) and of on-site pairs $n_k^{(2)}
\propto \sum_{\ell, \Delta} e^{-ik\Delta} \langle a^{\dagger}_{\ell+\Delta}a^{\dagger}_{\ell+\Delta}a_\ell a_\ell \rangle$ (blue, right axis), with $k=2\pi/Mn$, $n=0, \ldots, M-1$. Results are shown both for the exact ground state (solid lines) and the Ansatz (dashed lines). The inset shows the long-distance scaling of the correlation functions $\langle a^{\dagger}_{\ell+\Delta}a_\ell \rangle \sim \Delta^{-\alpha_1}$ (orange), and $\langle a^{\dagger}_{\ell+\Delta}a^{\dagger}_{\ell+\Delta}a_\ell a_\ell \rangle
\sim \Delta^{-\alpha_2}$ (blue), for the exact ground state (circles, $\alpha_1=0.22$, $\alpha_2=0.83$) and the Ansatz (diamonds, $\alpha_1=0.24$, $\alpha_2=0.99$. Parameters are $M=20$ and $\nu=1$.}
\end{figure}

Inspired by Cooper's ideas \cite{Cooper} for 2D rotating Bose gases we present an experimental scheme for the realization of Hamiltonian (\ref{Ham_3_body}).
Let us consider a system of bosonic atoms and diatomic Feshbach
molecules trapped in a 1D optical lattice. The
Hamiltonian of the system is $H=H_K+H_F+H_I$ \cite{Timmermans, Holland}, where 
\begin{eqnarray}
H_K&=&-t_a \sum_{i}(a_i^\dagger a_{i+1} + \textrm{h.c.}) -t_m \sum_{i} (m_i^\dagger m_{i+1} + \textrm{h.c.}),  \nonumber
\\
H_F&=&  \sum_{i} \Delta m_i^\dagger m_i + \frac{U_{aa} }{2}
  a_i^\dagger a_i^\dagger a_i a_i\nonumber  + \frac{g}{\sqrt{2} }  (m_i^\dagger a_i a_i + \textrm{h.c.}),
\\
 H_I &=& U_{am} \sum_{i=1}^M m_i^\dagger a_i^\dagger a_i m_i + \frac{U_{mm}}{2} \sum_{i=1}^M m_i^\dagger m_i^\dagger m_i m_i
\,  . \label{real} 
\end{eqnarray} 
Here, the bosonic operators for atoms
(molecules) $a_i$ ($m_i$) obey the usual canonical commutation
relations. The Hamiltonian $H_K$ describes the tunneling processes of atoms and molecules, occurring with amplitude $t_a$ and $t_m$, respectively. The term $H_F$ is the
Feshbach resonance term, with $\Delta$ being the energy off-set between
open and closed channels, $U_{aa}$ the on-site atom-atom
interaction and $g$ the coupling strength to the closed channel. Hamiltonian $H_I$ describes the on-site atom-molecule and molecule-molecule interactions.
We will assume a situation in which  $U_{aa}, U_{am},
U_{mm} \geq 0$, and $\Delta > 0$. Furthermore, we will consider the limit in which $\gamma^2=g^2/2\Delta^2 \ll 1$.
Within this limit the formation of molecules is highly suppressed due to the high energy offset, 
$\Delta$. However, virtual processes in which two atoms on the same lattice site go to the bound state, form a molecule and separate again, give rise to an effective 3-body interacting atomic Hamiltonian of the form \footnote{This effective Hamiltonian is obtained by projection of Hamiltonian (\ref{real}) onto the subspace with no molecules to first order in $\gamma^2$.}:
\begin{eqnarray} H_{\textrm{eff}}=&-&t_a
\sum_{i} (a_i^\dagger  a_{i+1} + \textrm{h.c.}) + U_{am}
\gamma^2 \sum_{i}(a_i^\dagger)^3 (a_i)^3 \nonumber \\&-& t_m
\gamma^2 \sum_{i}\left((a_i^\dagger)^2
(a_{i+1})^2 + \textrm{h.c.}\right) \nonumber \\
&+&(U_{aa}-g^2/\Delta)\sum_{i}(a_i^\dagger)^2 (a_i)^2, \label{effective}
\end{eqnarray}
where we have neglected higher order terms in $\gamma^2$.
Assuming $U_{aa}=g^2/\Delta$, and $t_m\gamma^2 \ll t_a$, $H_{\textrm{eff}}$ reduces to Hamiltonian (\ref{Ham}) with
$t=t_a$ and $U_3=U_{am}\gamma^2$. Finally, assuming  $U_{am}\gamma^2 \gg t_a $ we end up with the desired Hamiltonian (\ref{Ham_3_body}) for 3-hard-core bosons.

Let
us now summarize the requirements and approximations we have
imposed and discuss their experimental feasibility in typical setups with $^{87}$Rb. We have assumed $g^2/\Delta=U_{aa}$. Since $g=\sqrt{U_{aa}\Delta \mu \Delta B/2}$ \cite{Timmermans}, with $\Delta B$ being the width of the Feshbach resonance and $\Delta \mu$ the difference in magnetic momenta, we need $\Delta=\Delta \mu \Delta B/2$. For the Feshbach resonance at $1007.4 G$, this implies $\Delta/h=441$kHz \cite{Resonance, Durr1}. Furthermore, we have assumed $\gamma^2 \ll 1$, $t_m \gamma^2 \ll t_a$ and $U_{am} \gamma^2/t_a \gg 1$. Written in terms of the lattice and atomic and molecule parameters we have
$\gamma^2 =
\sqrt{32/\pi^3} (a_{aa}^{3D}a/a_\perp^2)  (\eta E_R / \Delta\mu
\Delta B)$,  $U_{am} / t_a =
(\sqrt{6}/4\pi)(a_{am}^{3D}a/a_\perp^2) \eta^{-2}
\exp(+\pi^2 \eta^2/4)$ and $t_m / t_a = 2 \exp(-\pi^2 \eta^2/4)$,
where $\eta=(V_0/E_R)^{1/4}$, with $V_0$ the lattice depth, $E_R=h^2/(8ma^2)$ the recoil energy, $m$ the atomic mass, and $a$ the lattice constant. The parameters $a_{aa}^{3D}$ and $a_{am}^{3D}$ are the 3D scattering length for atom-atom and atom-molecule collisions, and
$a_\perp = \sqrt{\hbar/\omega_\perp m}$ is the
transversal confinement width, with 
$\omega_\perp$ the transversal trapping frequency. Assuming typical values $\eta^4=50(70)$, $a_{aa}^{3D} \sim 5nm$ \cite{Durr2},
$a_{am}^{3D} \sim a_{aa}^{3D}$ \cite{Durr3},  $a=425nm$, and $\omega_\perp=2\pi \times 20$kHz \cite{Greiner}, we obtain:
$\gamma^2 = 3.55 \, (3.86) \times 10^{-3}$,
$t_m /t_a = 5.29 \, (0.22) \times 10^{-8}$, and $U_{am} \gamma^2 /
t_a = 1.35 \, (30.4) \times 10^3$, clearly satisfying the required conditions.

Regarding detection of the Pfaffian-like ground state, the characteristic difference between both the momentum distribution and number statistics of particles and on-site pairs could be observed via spin-changing collisions \cite{Widera}.

In conclusion, we have shown that the ground state of 3-hard-core bosons in a 1D lattice can be well described by a Pfaffian-like state which is a cluster of two T-G gases. We have shown that such a state may be accessible with current technology with atoms and molecules in optical lattices. We believe that our findings may open a new path for the creation of NAA.

B. Paredes would like to thank M. Greiter for estimulating discussions and critical reading of this manuscript.

\end{document}